\documentclass{article}

\usepackage[preprint]{neurips_2021}

\usepackage[utf8]{inputenc} % allow utf-8 input
\usepackage[T1]{fontenc}    % use 8-bit T1 fonts
\usepackage{hyperref}       % hyperlinks
\usepackage{url}            % simple URL typesetting
\usepackage{booktabs}       % professional-quality tables
\usepackage{amsfonts}       % blackboard math symbols
\usepackage{nicefrac}       % compact symbols for 1/2, etc.
\usepackage{microtype}      % microtypography
\usepackage{xcolor}         % colors
\usepackage{algpseudocode}  % algorithm
\usepackage{algorithm}
\usepackage{graphicx}       % display figure
\usepackage{amsmath}
\title{End to End Generative Meta Curriculum Learning For Medical Data Augmentation}

\author{%
  Meng Li\\
  School of Information Technology and Electrical Engineering\\
  The University of Queensland, \\
  St Lucia, QLD 4072, Australia \\
  \texttt{meng.li@uq.edu.au} \\
}

\begin{document}

\maketitle

\begin{abstract}
Current medical image synthetic augmentation techniques rely on intensive use of generative adversarial networks (GANs). However, the nature of GAN architecture leads to heavy computational resources to produce synthetic images and the augmentation process requires multiple stages to complete. To address these challenges, we introduce a novel generative meta curriculum learning method that trains the task-specific model (student) end-to-end with only one additional teacher model. The teacher learns to generate curriculum to feed into the student model for data augmentation and guides the student to improve performance in a meta-learning style. In contrast to the generator and discriminator in GAN, which compete with each other, the teacher and student collaborate to improve the student's performance on the target tasks. Extensive experiments on the histopathology datasets show that leveraging our framework results in significant and consistent improvements in classification performance.
\end{abstract}

\section{Introduction}

Existing deep learning methods have achieved decent performance on histopathological image analysis tasks. However, this requires abundantly labeled data and most of works come at the expense of acquisition of detailed labels and extensive participation of experts~\cite{nie2017medical}. The whole process is labor-intensive and time-consuming, which is impractical for rare diseases, early clinical studies, or new imaging modalities~\cite{medela2019few}. Recently, some works have addressed this issue by proposing image synthesis methods for data augmentation and achieved satisfactory performance~\cite{nie2017medical,frid2018synthetic,dar2019image}. These works commonly employed GAN architectures~\cite{huang2018empirical} for image generation. The data augmentation of these works takes two stage: 1) generating synthetic images, and 2) combining these images with original training data for data augmentation. The framework includes at least three individual networks: a generator, a discriminator, and a task-specific network. This is cumbersome, which needs expensive computing resources for GANs to generate images and requires more effort to train the networks. Traditionally, different tasks require specific type of synthetic images for data augmentation, a more general data augmentation approach that does not require specific training image features will undoubtedly improve the efficiency and benefit all tasks.

We address the aforementioned problems by introducing a novel generative teacher-student meta curriculum learning framework (GMCL) for histopathology image data augmentation. This is different to generator-discriminator in GAN that compete with each other, the teacher and student in GMCL collaborate to improve the student performance on the target tasks. In this framework, only two individual networks are needed: a generator (teacher) and a task-specific network (student). In addition, The teacher does not need have knowledge about the features of training data, which means that this framework can handle any specific task. The data augmentation process in GMCL is totally end-to-end and the training process is elegant and more efficient.

\section{Related Work}
\label{related_work}
\subsection{Teacher-Student Models}
Modelling deep learning networks as a teacher-student setting is common in recent works. One prominent employment is model compression by knowledge distillation~\cite{gou2021knowledge, hinton2015distilling}. The main objective in knowledge distillation is to transfer the knowledge from a large teacher model to a small student model. Earlier work trains a student network through the softmax prediction of a teacher model~\cite{hinton2015distilling}. Instead of using soft labels from the teacher, methods that manually design soft labels are proposed, such as label smoothing~\cite{muller2019does} and temperature dampening~\cite{xie2020unsupervised}. Beyond the design of soft labels, the student model performance is employed as the feedback to update teacher model~\cite{pham2021meta}. Besides knowledge distillation, the teacher-student architecture is implemented as reinforcement learning (RL) agents for RL tasks~\cite{zimmer2014teacher}. It can also be used to perform conditional unsupervised domain adaptation on acoustic data~\cite{meng2018adversarial}. In our teacher-student framework, the teacher network is implemented as a generative model, and the output is fed into the student for learning. The student model's performance is used as the feedback to update the teacher.

\subsection{Bi-Level Optimization Algorithms}
In our framework, the update rule of the teacher network is derived from the student's feedback. This method is based on a bi-level optimization algorithm that commonly appears in meta-learning literature. There are different problems that bi-level optimization methods try to optimize, such as re-weighting training data~\cite{ren2018learning, ren2020not}, differentiable architecture search~\cite{liu2018darts}, learning rate adaptation~\cite{baydin2017online}, training label correction~\cite{zheng2019meta}, and generative learning optimization~\cite{such2020generative}. Our model adopts the similar bi-level optimization technique to these works. The difference between our work and these methods is that we implement the bi-level optimization algorithm to optimize generated augmentation data from the teacher to boost the student performance.

\subsection{Data Augmentation}
Data augmentation techniques have been broadly used by deep learning networks to reduce over-fitting and improve generalization performance. In traditional literature, data augmentation is implemented based on basic image manipulations, such as geometric transformations, flipping, rotation, and cropping~\cite{shorten2019survey}. Data augmentation can also be achieved through deep learning. Feature space augmentation transforms higher dimension features into lower representations~\cite{devries2017dataset}. Adversarial training frameworks train a rival network that learns augmentations which encourage misclassification in its rival classification network~\cite{moosavi2016deepfool}. GAN-based models generate synthesized images, which can also be used as augmentation data to improve task performance~\cite{mirza2014conditional, yi2019generative}. Meta learning techniques have also been implemented to learn data augmentation policy, such as a meta-learning neural style augmentation~\cite{perez2017effectiveness}, and a reinforcement learning strategy that explores an optimal policy amongst a set of geometric transformations~\cite{cubuk2018autoaugment}. We propose a meta-learning method that trains a teacher model to generate augmentation data based on the feedback from the student network.

~\cite{yi2019generative}
\subsection{Medical Image Synthetic Augmentation}
Traditionally, medical image synthesis is achieved by working on the texture features, such as decomposing RGB color image of skin into texture components~\cite{doi2006image}, or modeling synthesis transformation as a nonlinear regression of image patches~\cite{jog2013magnetic}. In recent works, GANs are commonly used. A context-aware GAN model~\cite{nie2017medical} proposed a solution to generate computed tomography (CT) images based on magnetic resonance images. ~\cite{frid2018synthetic} used a traditional GAN to generate liver lesions CT images for data augmentation. To generate multi-contrast synthesized images, ~\cite{dar2019image} employed a conditional GAN with perceptual and cycle-consistency losses to preserve intermediate-to-high frequency details to enhance MRI generation performance. To achieve better medical image tumor segmentation accuracy, ~\cite{zhang2017deep} proposed a consistent evaluation GAN-based network. ~\cite{xue2019synthetic} implemented a data augmentation method, by filtering synthetic-image based on the divergence in feature space between synthesized images and class centroids. Recently, transformer-based GANs for medical images were explored and proposed an EncoderCNN-Transformer-DecoderCNN network~\cite{luo20213d}. Particularly, the transformer module is targeted at capturing the long-range dependencies between the features learned by the EncoderCNN. In this work, our teacher model serves as a generative model for image generation. Our teacher and student are trained collaboratively, as opposed to adversarially in GANs.

\section{Method}

\subsection{End-to-End Data Augmentation}
Data augmentation refers to the generation of more data from existing data to augment the training set. The generator in a GAN performs data augmentation by synthesizing data as extra training data. The proposed GMCL also has two networks: the teacher and the student. However, GMCL differs from previous data augmentation approaches in that:
1) the teacher and student are trained collaboratively, as opposed to adversarially in GANs. 2) the teacher and student are trained together, in contrast to training and freezing the generator, then training the discriminator in GANs. 3) The generated data (curriculum) does not necessarily look realistic as the information it contains is more than just image features.

\begin{figure}[t]
  \centering
  \includegraphics[height=3.7cm,keepaspectratio]{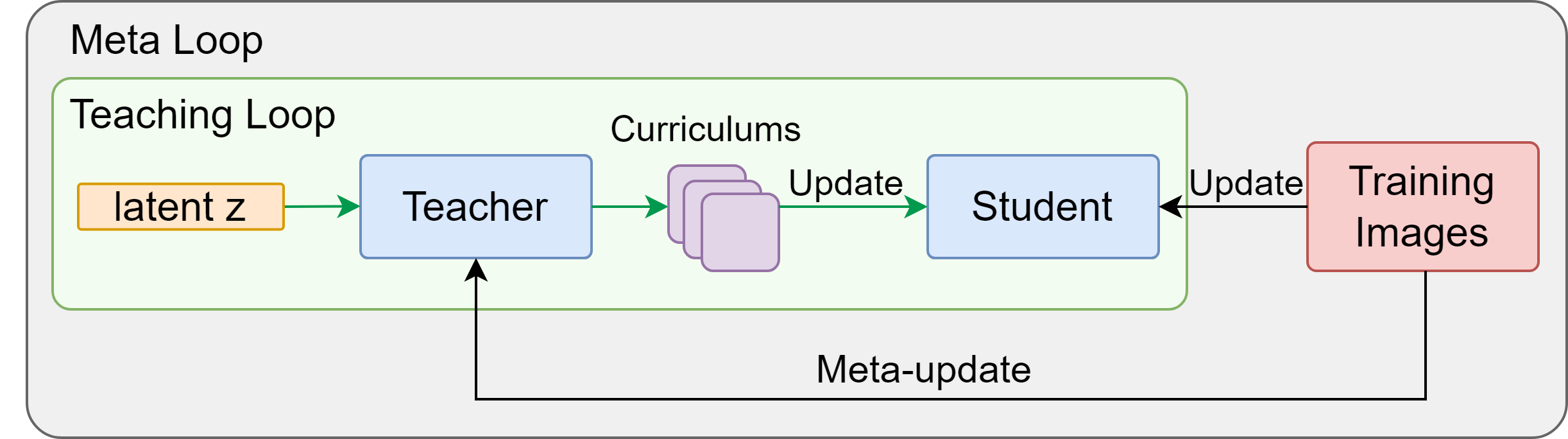}
  \caption{Overview of our proposed GMCL network. In teaching loop, the conditioned latent vector $z$ is fed to the Teacher and produces a batch of curriculum. The Student network takes the curriculum as the input and makes prediction. The loss is used to update the Student. In meta loop, the real training images are evaluated by the trained Student. The loss is used to update parameters of both Student and Teacher networks.}
  \label{fig:net_arch}
\end{figure}

\subsection{Notations}
Let $T$ and $S$ denote the teacher network and the student network respectively. Let the teacher parameters be $\theta_{T}$ and $\theta_{meta}$, and the student parameters be $\theta_{S}$. We use $\left(X_{i}, Y_{i}\right)$ to denote a batch of real images and their corresponding labels. We use $\left(Z_{j}, Y_{j}\right)$ to refer to a batch of noise vectors and their corresponding labels. We denote by $T\left(Z_{j} , Y_{j} ; \theta_{T}\right)$ the curriculum generated by the teacher. Note that the dimension of the curriculum is same as the real image. The prediction of the student is denoted as $S\left(X_{i} ; \theta_{S}\right)$. We use $\mathrm{CE}(p, q)$ to refer to the cross-entropy loss between two distributions p and q. The learning rate of the student and the teacher is denoted as $\alpha$ and $\beta$.

\subsection{Teaching Loop}
The overall pipeline of our proposed end-to-end GMCL network is shown in Fig.~\ref{fig:net_arch}. Unlike the generator and the discriminator in GAN that compete with each other, the teacher and the student in GMCL collaborate to improve the student performance on the target tasks. There are two nested loops in the network. In the inner teaching loop, only the student is optimized, the objective is given by:

\begin{equation} \label{eq:1}
\theta_{S}^{\mathrm{teach}}=\underset{\theta_{S}}{\operatorname{argmin}} \; \; 
\underbrace{
\mathbb{E}\left[\mathrm{CE}\left(S\left(T\left(Z_{j}, Y_{j} ; \theta_{T}\right) ; \theta_{S}\right), Y_{j} \right)\right]
}_{:=\mathcal{L}_{teach}\left(\theta_{T}, \theta_{S}\right)},
\end{equation}

where the curriculum $T\left(Z_{j}, Y_{j} ; \theta_{T}\right)$ is produced by a conditioned teacher that is trained in outer meta loop with fixed parameters $\theta_{T}$. The curriculum is of the same dimension as the real image, and then fed to the student for training a predefined number of teaching iterations. The student network is optimized based on objective~\ref{eq:1} with SGD: 
$\theta_{S}^{\prime}=\theta_{S} - \alpha \nabla_{\theta_{S}} \mathcal{L}_{teach}\left(\theta_{T}, \theta_{S}\right)$.
In this work, different strategies to implement $Z_{j}$ are compared in Section \ref{experiments}. In this work, we use fixed $Z_{j}$ for each outer meta loop.
$ Here we can try to produce both image and label ??? $

\subsection{Meta Loop}
In one outer meta loop iteration, both the student and teacher network are optimized. A random batch of training data $\left(X_{i}, Y_{i}\right)$ is firstly sampled, and then evaluated on the student that is previously trained by teacher's curriculum. The objective is given by:

\begin{equation} \label{eq:2}
\theta_{S}^{\mathrm{meta}}=\underset{\theta_{S}}{\operatorname{argmin}} \; \; 
\underbrace{
\mathbb{E}\left[\mathrm{CE}\left(S\left(X_{i} ; \theta_{S}\right), Y_{i} \right)\right]
}_{:=\mathcal{L}_{meta}\left(\theta_{S}\right)}.
\end{equation}

In addition to learning from the training data, the student network is also trained using curriculum information from the teacher, which can be regarded as the \emph{data augmentation} process. We then optimize the teacher parameters on a meta level, by using the loss $\mathcal{L}_{meta}$ of student performance, which leads to the practical teacher objective: 

\begin{equation} \label{eq:3}
\min_{\theta_{T}} \; \; \mathcal{L}_{meta}(\underbrace{\theta_{S} - \alpha \nabla_{\theta_{S}} \mathcal{L}_{teach}\left(\theta_{T}. \theta_{S}\right)}_{=\theta_{S}^{\prime} \text { reused from teaching loop update}}).
\end{equation}

In the meta loop, the student is optimized using real data based on objective~\ref{eq:2} with SGD: 
$\theta_{S}^{\prime\prime}=\theta_{S} - \alpha \nabla_{\theta_{S}} \mathcal{L}_{meta}\left(\theta_{S}^{\prime}\right)$. 
On the other hand, the teacher is optimized using the feedback from the student based on objective~\ref{eq:3} with Adam:
$\theta_{T}^{\prime}=\theta_{T}- \beta \nabla_{\theta_{T}} \mathcal{L}_{meta}(\theta_{S}^{\prime})$. 

The teacher is not only trained to generate curriculum that instructs the student to learn how to classify, it also provides guidance on appropriate learning rate and momentum values by employing meta parameters. These parameters are also updated based on the same loss by student on training data, with a different learning rate. More details are in Algorithm \ref{alg:GMCL}.

\begin{algorithm}
\caption{Generative Meta Curriculum Learning}\label{alg:GMCL}
\begin{algorithmic}[1]
\renewcommand{\algorithmicrequire}{\textbf{Input:}}
\Require Training set $D_{train}$, teacher $T$ with parameters $\theta_{T}$ and meta parameters $\theta_{meta}$, student $S$ with parameters $\theta_{S}$, step sizes $\alpha$, $\beta$, $\beta_{meta}$ for student, teacher, and meta parameters of teacher separately.
\For{$i=1,N$} \Comment{Meta loop}
    \For{$j=1,M$} \Comment{Teaching loop}
        \State $\mathbf{Z}_{j} \sim p(z)$ one batch of prior samples
        \State $\mathbf{O}_{j} = T(\mathbf{Z}_{j},{Y}_{j}  ; \theta_{T})$ \Comment{Generate conditional curriculum}
        \State $\mathbf{C}_{j} = S(\mathbf{O}_{j} ; \theta_{S})$
        \State $g_{\theta_{S}} \gets \nabla_{\theta_{S}}($CE$ (C_{j}, Y_{j}))$ \Comment{Calculate gradient based on cross-entropy loss}
        \State $\alpha_{j}, \mu_{j} \gets \theta_{meta}$ \Comment{Set step size and momentum to meta parameters}
        \State $\theta_{S} \gets \theta_{S} - \alpha_{j} \cdot$ SGD$(\theta_{S}, g_{\theta_{S}},\mu_{j})$ \Comment{Update student}
    \EndFor
    \State $\mathbf{X}_{i}, \mathbf{Y}_{i} \gets$ random batch from $D_{train}$
    \State $\mathbf{C}_{i} = S(\mathbf{X}_{i} ; \theta_{S})$
    
    \State $g_{\theta_{T}} \gets \nabla_{\theta_{T}}($CE$ (C_{i}, Y_{i}))$
    \State $g_{\theta_{S}} \gets \nabla_{\theta_{S}}($CE$ (C_{i}, Y_{i}))$
    
    \State $\theta_{S} \gets \theta_{S} - \alpha_{M} \cdot$ SGD$(\theta_{S}, g_{\theta_{S}},\mu_{M})$  \Comment{Update student}    
    \State $\theta_{T} \gets \theta_{T} - \beta \cdot$ ADAM$(\theta_{T}, g_{\theta_{T}})$  \Comment{Update teacher}
    \State $\theta_{meta} \gets \theta_{meta} - \beta_{meta} \cdot$ ADAM$(\theta_{meta}, g_{\theta_{T}})$ \Comment{Update meta parameters of teacher}
    
\EndFor
\end{algorithmic}
\end{algorithm}

\section{Experiments}
\label{experiments}
\subsection{Datasets}
We conduct experiments on two public datasets: colorectal adenocarcinoma (Colon Cancer) dataset~\cite{sirinukunwattana2016locality} and Patc hCamelyon (PCam) benchmark dataset\cite{veeling2018rotation}. The Colon Cancer dataset contains 29,756 color images acquired and digitized at a resolution of $0.55\mu m/pixel$. All images are manually annotated by experienced pathologists. We divided the image into 32 $\times$ 32 patches and split into four classes, \textit{i.e.} epithelial, inflammatory, fibroblast, and miscellaneous.

\subsection{Implementation Details}
The GMCL network architecture is a small CNN network. The generator contains two fully connected layers with $1024$ and $128 \times H/4 \times H/4$ filters, respectively, where H is the dimension of the synthetic image. Then, there are 2 convolutional layers after fully connected layers. The first convolutional layer has 64 filters. The second convolutional layer has 1 filter followed by a Tanh activation function. We set stride=2 for the second and fourth convolution layers for dimension reduction. The learner contains 5 convolutional layers followed by a global average pooling and an FC layer. We set stride=2 for the second and fourth convolution layers for dimension reduction. We augment the real training images with random crops and horizontal flips. We used Kaiming Normal initialization and LeakyReLUs (with $\alpha$ = 0.1)~\cite{he2015delving}. We use Batch normalization (BatchNorm) for both the generator and the learner. The BatchNorm momentum for the learner was set to 0. More hyper-parameters settings can be found in Table~\ref{sec8:hyperparameters}. All experiments were conducted on two Tesla V100 GPUs, each with 16G of RAM.

\begin{table}[t]
	\centering
	\caption{Hyperparameters for experiments}
	\label{sec8:hyperparameters}
	\begin{tabular}{|c|c|}
		\hline Hyperparameter & Value \\
		\hline \hline Learning Rate & $0.02$ \\
		\hline Initial LR & $0.02$ \\
		\hline Initial Momentum & $0.5$ \\
		\hline Adam Beta\textunderscore1 & $0.9$ \\
		\hline Adam Beta\textunderscore2 & $0.9$ \\
		\hline Adam $\epsilon$ & $1 \mathrm{e}-5$ \\
		\hline Size of latent variable & 128 \\
		\hline Inner-loop Batch Size & 64 \\
		\hline Outer-loop Batch Size & 128 \\
		\hline
	\end{tabular}
\end{table}

\subsection{Results and Discussions}
\label{sec8:results}
We conduct experiments on different methods and the results are shown in Table~\ref{sec8:Comparison}. 
As the generative meta curriculum learning for data augmentation is less studied in previous literature. We compare our method with the plain learner (without teacher network), and the generative teaching network (GTN) for histopathology image classification task. 
Different evaluation metrics are selected for comparison. They are accuracy, the area under the ROC curve (AUC), sensitivity, and specificity for a comprehensive comparison. We also compare our GMCL augmentation method with traditional augmentation techniques (random rotate, random flip, and color jitter). The results consistently demonstrate that our GMCL method achieve better performance.

\begin{figure}[t]
	\centering
	\includegraphics[height=4.7cm,keepaspectratio]{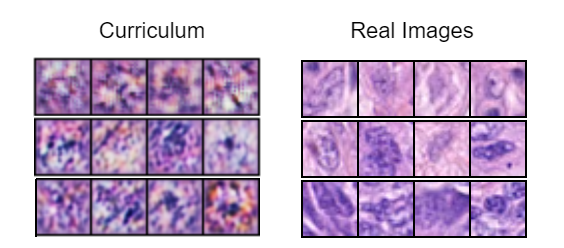}
	\caption{Generated curriculum and the real images.}
	\label{fig:GMCL_gen_imgs}
\end{figure}

We visualize the curriculum generated by the teacher network in Figure~\ref{fig:GMCL_gen_imgs}. Unlike the synthetic images for augmentation in other methods, the visualized curriculum does not look completely similar to the real images. The reason is that the curriculum includes more information than the image itself, it also contains the instructions for student to adapt its hyperparameters.

\begin{table}[t]
	\centering
	\caption{Comparison of different methods for Colon Cancer dataset.}
	\label{sec8:Comparison}
	\begin{tabular}{c|c|c|c|c} 
		\hline
		Method                  & Acc             & \multicolumn{1}{l|}{AUC} & \multicolumn{1}{l|}{Sensitivity} & \multicolumn{1}{l}{Specificity}  \\ 
		\hline
		Plain Learner           & 0.6762          & 0.8812                   & 0.8723                           & 0.9211                           \\ 
		\hline
		Plain Learner + Aug     & 0.6912          & 0.8991                   & 0.8735                           & 0.9132                           \\ 
		\hline
		GTN                     & 0.7314          & 0.9176                   & 0.8812                           & 0.9190                           \\ 
		\hline
		GMCL (w/o adaptive LR)  & 0.7421          & 0.9173                   & 0.8920                           & 0.9205                           \\ 
		\hline
		GMCL (w/ adaptive LR) & \textbf{0.7531} & \textbf{0.9200}          & \textbf{0.8987}                  & \textbf{0.9241}                  \\
		\hline
	\end{tabular}							
\end{table}

\section{Conclusions}
In this work, we address the issue that current histopathology data augmentation methods are less efficient and cumbersome, by proposing an end-to-end generative teacher-student meta curriculum learning framework for histopathology image data augmentation. Experiment results show that our method demonstrates better performance than other methods.

% \begin{ack}
% This project has been funded by Sullivan Nicolaides Pathology and the Australian Research Council (ARC) Linkage Project [Grant number LP160101797].
% \end{ack}

\small

\bibliography{bib.bib}

\begin{thebibliography}{33}
\providecommand{\natexlab}[1]{#1}
\providecommand{\url}[1]{\texttt{#1}}
\expandafter\ifx\csname urlstyle\endcsname\relax
  \providecommand{\doi}[1]{doi: #1}\else
  \providecommand{\doi}{doi: \begingroup \urlstyle{rm}\Url}\fi

\bibitem[Baydin et~al.(2017)Baydin, Cornish, Rubio, Schmidt, and
  Wood]{baydin2017online}
A.~G. Baydin, R.~Cornish, D.~M. Rubio, M.~Schmidt, and F.~Wood.
\newblock Online learning rate adaptation with hypergradient descent.
\newblock \emph{arXiv preprint arXiv:1703.04782}, 2017.

\bibitem[Cubuk et~al.(2018)Cubuk, Zoph, Mane, Vasudevan, and
  Le]{cubuk2018autoaugment}
E.~D. Cubuk, B.~Zoph, D.~Mane, V.~Vasudevan, and Q.~V. Le.
\newblock Autoaugment: Learning augmentation policies from data.
\newblock \emph{arXiv preprint arXiv:1805.09501}, 2018.

\bibitem[Dar et~al.(2019)Dar, Yurt, Karacan, Erdem, Erdem, and
  {\c{C}}ukur]{dar2019image}
S.~U. Dar, M.~Yurt, L.~Karacan, A.~Erdem, E.~Erdem, and T.~{\c{C}}ukur.
\newblock Image synthesis in multi-contrast mri with conditional generative
  adversarial networks.
\newblock \emph{IEEE transactions on medical imaging}, 38\penalty0
  (10):\penalty0 2375--2388, 2019.

\bibitem[DeVries and Taylor(2017)]{devries2017dataset}
T.~DeVries and G.~W. Taylor.
\newblock Dataset augmentation in feature space.
\newblock \emph{arXiv preprint arXiv:1702.05538}, 2017.

\bibitem[Doi and Tominaga(2006)]{doi2006image}
M.~Doi and S.~Tominaga.
\newblock Image analysis and synthesis of skin color textures by wavelet
  transform.
\newblock In \emph{2006 IEEE Southwest Symposium on Image Analysis and
  Interpretation}, pages 193--197. IEEE, 2006.

\bibitem[Frid-Adar et~al.(2018)Frid-Adar, Klang, Amitai, Goldberger, and
  Greenspan]{frid2018synthetic}
M.~Frid-Adar, E.~Klang, M.~Amitai, J.~Goldberger, and H.~Greenspan.
\newblock Synthetic data augmentation using gan for improved liver lesion
  classification.
\newblock In \emph{2018 IEEE 15th international symposium on biomedical imaging
  (ISBI 2018)}, pages 289--293. IEEE, 2018.

\bibitem[Gou et~al.(2021)Gou, Yu, Maybank, and Tao]{gou2021knowledge}
J.~Gou, B.~Yu, S.~J. Maybank, and D.~Tao.
\newblock Knowledge distillation: A survey.
\newblock \emph{International Journal of Computer Vision}, 129\penalty0
  (6):\penalty0 1789--1819, 2021.

\bibitem[He et~al.(2015)He, Zhang, Ren, and Sun]{he2015delving}
K.~He, X.~Zhang, S.~Ren, and J.~Sun.
\newblock Delving deep into rectifiers: Surpassing human-level performance on
  imagenet classification.
\newblock In \emph{Proceedings of the IEEE international conference on computer
  vision}, pages 1026--1034, 2015.

\bibitem[Hinton et~al.(2015)Hinton, Vinyals, Dean,
  et~al.]{hinton2015distilling}
G.~Hinton, O.~Vinyals, J.~Dean, et~al.
\newblock Distilling the knowledge in a neural network.
\newblock \emph{arXiv preprint arXiv:1503.02531}, 2\penalty0 (7), 2015.

\bibitem[Huang et~al.(2018)Huang, Yuan, Xu, Guo, Sun, Wu, and
  Weinberger]{huang2018empirical}
G.~Huang, Y.~Yuan, Q.~Xu, C.~Guo, Y.~Sun, F.~Wu, and K.~Weinberger.
\newblock An empirical study on evaluation metrics of generative adversarial
  networks.
\newblock 2018.

\bibitem[Jog et~al.(2013)Jog, Roy, Carass, and Prince]{jog2013magnetic}
A.~Jog, S.~Roy, A.~Carass, and J.~L. Prince.
\newblock Magnetic resonance image synthesis through patch regression.
\newblock In \emph{2013 IEEE 10th International Symposium on Biomedical
  Imaging}, pages 350--353. IEEE, 2013.

\bibitem[Liu et~al.(2018)Liu, Simonyan, and Yang]{liu2018darts}
H.~Liu, K.~Simonyan, and Y.~Yang.
\newblock Darts: Differentiable architecture search.
\newblock \emph{arXiv preprint arXiv:1806.09055}, 2018.

\bibitem[Luo et~al.(2021)Luo, Wang, Zu, Zhan, Wu, Zhou, Shen, and
  Zhou]{luo20213d}
Y.~Luo, Y.~Wang, C.~Zu, B.~Zhan, X.~Wu, J.~Zhou, D.~Shen, and L.~Zhou.
\newblock 3d transformer-gan for high-quality pet reconstruction.
\newblock In \emph{International Conference on Medical Image Computing and
  Computer-Assisted Intervention}, pages 276--285. Springer, 2021.

\bibitem[Medela et~al.(2019)Medela, Picon, Saratxaga, Belar, Cabez{\'o}n,
  Cicchi, Bilbao, and Glover]{medela2019few}
A.~Medela, A.~Picon, C.~L. Saratxaga, O.~Belar, V.~Cabez{\'o}n, R.~Cicchi,
  R.~Bilbao, and B.~Glover.
\newblock Few shot learning in histopathological images: reducing the need of
  labeled data on biological datasets.
\newblock In \emph{2019 IEEE 16th International Symposium on Biomedical Imaging
  (ISBI 2019)}, pages 1860--1864. IEEE, 2019.

\bibitem[Meng et~al.(2018)Meng, Li, Gong, and Juang]{meng2018adversarial}
Z.~Meng, J.~Li, Y.~Gong, and B.-H. Juang.
\newblock Adversarial teacher-student learning for unsupervised domain
  adaptation.
\newblock In \emph{2018 IEEE International Conference on Acoustics, Speech and
  Signal Processing (ICASSP)}, pages 5949--5953. IEEE, 2018.

\bibitem[Mirza and Osindero(2014)]{mirza2014conditional}
M.~Mirza and S.~Osindero.
\newblock Conditional generative adversarial nets.
\newblock \emph{arXiv preprint arXiv:1411.1784}, 2014.

\bibitem[Moosavi-Dezfooli et~al.(2016)Moosavi-Dezfooli, Fawzi, and
  Frossard]{moosavi2016deepfool}
S.-M. Moosavi-Dezfooli, A.~Fawzi, and P.~Frossard.
\newblock Deepfool: a simple and accurate method to fool deep neural networks.
\newblock In \emph{Proceedings of the IEEE conference on computer vision and
  pattern recognition}, pages 2574--2582, 2016.

\bibitem[M{\"u}ller et~al.(2019)M{\"u}ller, Kornblith, and
  Hinton]{muller2019does}
R.~M{\"u}ller, S.~Kornblith, and G.~E. Hinton.
\newblock When does label smoothing help?
\newblock \emph{Advances in neural information processing systems}, 32, 2019.

\bibitem[Nie et~al.(2017)Nie, Trullo, Lian, Petitjean, Ruan, Wang, and
  Shen]{nie2017medical}
D.~Nie, R.~Trullo, J.~Lian, C.~Petitjean, S.~Ruan, Q.~Wang, and D.~Shen.
\newblock Medical image synthesis with context-aware generative adversarial
  networks.
\newblock In \emph{International conference on medical image computing and
  computer-assisted intervention}, pages 417--425. Springer, 2017.

\bibitem[Perez and Wang(2017)]{perez2017effectiveness}
L.~Perez and J.~Wang.
\newblock The effectiveness of data augmentation in image classification using
  deep learning.
\newblock \emph{arXiv preprint arXiv:1712.04621}, 2017.

\bibitem[Pham et~al.(2021)Pham, Dai, Xie, and Le]{pham2021meta}
H.~Pham, Z.~Dai, Q.~Xie, and Q.~V. Le.
\newblock Meta pseudo labels.
\newblock In \emph{Proceedings of the IEEE/CVF Conference on Computer Vision
  and Pattern Recognition}, pages 11557--11568, 2021.

\bibitem[Ren et~al.(2018)Ren, Zeng, Yang, and Urtasun]{ren2018learning}
M.~Ren, W.~Zeng, B.~Yang, and R.~Urtasun.
\newblock Learning to reweight examples for robust deep learning.
\newblock In \emph{International conference on machine learning}, pages
  4334--4343. PMLR, 2018.

\bibitem[Ren et~al.(2020)Ren, Yeh, and Schwing]{ren2020not}
Z.~Ren, R.~Yeh, and A.~Schwing.
\newblock Not all unlabeled data are equal: Learning to weight data in
  semi-supervised learning.
\newblock \emph{Advances in Neural Information Processing Systems},
  33:\penalty0 21786--21797, 2020.

\bibitem[Shorten and Khoshgoftaar(2019)]{shorten2019survey}
C.~Shorten and T.~M. Khoshgoftaar.
\newblock A survey on image data augmentation for deep learning.
\newblock \emph{Journal of big data}, 6\penalty0 (1):\penalty0 1--48, 2019.

\bibitem[Sirinukunwattana et~al.(2016)Sirinukunwattana, Raza, Tsang, Snead,
  Cree, and Rajpoot]{sirinukunwattana2016locality}
K.~Sirinukunwattana, S.~E.~A. Raza, Y.-W. Tsang, D.~R. Snead, I.~A. Cree, and
  N.~M. Rajpoot.
\newblock Locality sensitive deep learning for detection and classification of
  nuclei in routine colon cancer histology images.
\newblock \emph{IEEE transactions on medical imaging}, 35\penalty0
  (5):\penalty0 1196--1206, 2016.

\bibitem[Such et~al.(2020)Such, Rawal, Lehman, Stanley, and
  Clune]{such2020generative}
F.~P. Such, A.~Rawal, J.~Lehman, K.~Stanley, and J.~Clune.
\newblock Generative teaching networks: Accelerating neural architecture search
  by learning to generate synthetic training data.
\newblock In \emph{International Conference on Machine Learning}, pages
  9206--9216. PMLR, 2020.

\bibitem[Veeling et~al.(2018)Veeling, Linmans, Winkens, Cohen, and
  Welling]{veeling2018rotation}
B.~S. Veeling, J.~Linmans, J.~Winkens, T.~Cohen, and M.~Welling.
\newblock Rotation equivariant cnns for digital pathology.
\newblock In \emph{International Conference on Medical image computing and
  computer-assisted intervention}, pages 210--218. Springer, 2018.

\bibitem[Xie et~al.(2020)Xie, Dai, Hovy, Luong, and Le]{xie2020unsupervised}
Q.~Xie, Z.~Dai, E.~Hovy, T.~Luong, and Q.~Le.
\newblock Unsupervised data augmentation for consistency training.
\newblock \emph{Advances in Neural Information Processing Systems},
  33:\penalty0 6256--6268, 2020.

\bibitem[Xue et~al.(2019)Xue, Zhou, Ye, Long, Antani, Cornwell, Xue, and
  Huang]{xue2019synthetic}
Y.~Xue, Q.~Zhou, J.~Ye, L.~R. Long, S.~Antani, C.~Cornwell, Z.~Xue, and
  X.~Huang.
\newblock Synthetic augmentation and feature-based filtering for improved
  cervical histopathology image classification.
\newblock In \emph{International conference on medical image computing and
  computer-assisted intervention}, pages 387--396. Springer, 2019.

\bibitem[Yi et~al.(2019)Yi, Walia, and Babyn]{yi2019generative}
X.~Yi, E.~Walia, and P.~Babyn.
\newblock Generative adversarial network in medical imaging: A review.
\newblock \emph{Medical image analysis}, 58:\penalty0 101552, 2019.

\bibitem[Zhang et~al.(2017)Zhang, Yang, Chen, Fredericksen, Hughes, and
  Chen]{zhang2017deep}
Y.~Zhang, L.~Yang, J.~Chen, M.~Fredericksen, D.~P. Hughes, and D.~Z. Chen.
\newblock Deep adversarial networks for biomedical image segmentation utilizing
  unannotated images.
\newblock In \emph{International conference on medical image computing and
  computer-assisted intervention}, pages 408--416. Springer, 2017.

\bibitem[Zheng et~al.(2019)Zheng, Awadallah, and Dumais]{zheng2019meta}
G.~Zheng, A.~H. Awadallah, and S.~Dumais.
\newblock Meta label correction for learning with weak supervision.
\newblock \emph{arXiv preprint arXiv:1911.03809}, 2019.

\bibitem[Zimmer et~al.(2014)Zimmer, Viappiani, and Weng]{zimmer2014teacher}
M.~Zimmer, P.~Viappiani, and P.~Weng.
\newblock Teacher-student framework: a reinforcement learning approach.
\newblock In \emph{AAMAS Workshop Autonomous Robots and Multirobot Systems},
  2014.

\end{thebibliography}

% [1] Alexander, J.A.\ \& Mozer, M.C.\ (1995) Template-based algorithms for
% connectionist rule extraction. In G.\ Tesauro, D.S.\ Touretzky and T.K.\ Leen
% (eds.), {\it Advances in Neural Information Processing Systems 7},
% pp.\ 609--616. Cambridge, MA: MIT Press.

\end{document}